\documentclass[prd,aps,showpacs,epsf,amsfonts,floats,amssymb,amsmath,nofootinbib]{jpconf}

\newcommand{\be}{\begin{equation}}\newcommand{\ee}{\end{equation}}
\newcommand{\bea}{\begin{eqnarray}}\newcommand{\eea}{\end{eqnarray}}
\newcommand{\beaa}{\begin{eqnarray}}\newcommand{\eeaa}{\end{eqnarray}}
\newcommand{\ba}{\begin{array}}\newcommand{\ea}{\end{array}}
\newcommand{\bit}{\begin{itemize}}\newcommand{\eit}{\end{itemize}}
\newcommand{\ben}{\begin{enumerate}}\newcommand{\een}{\end{enumerate}}

\def\pa{\partial}
\def\rar{\rightarrow}

\def\om{\omega}

\def\1{{_{1}}}\def\2{{_{2}}}
\def\ZzZ{{\hbox{\tenrm Z\kern-.31em{Z}}}}
\def\CcC{{\hbox{\tenrm C\kern-.45em{\vrule height.67em width0.08em depth-
.04em \hskip.45em }}}}

\newcommand{\lab}{\label}

\newcommand{\bc}{\begin{center}}
\newcommand{\ec}{\end{center}}
\begin{document}
%
%%\preprint{salerno dmi}
\title{Geometric phase and gauge theory structure\\
 in quantum computing}
\author{A. Bruno${}^{1,3}$,  A. Capolupo${}^{2,3}$, S. Kak${}^4$, G. Raimondo${}^{1,3}$
and G. Vitiello${}^{2,3}$}
\address{${}^{1}$Dipartimento di Fisica dell'Universit\`a di Salerno, 84100 Salerno, Italy}

\address{${}^{2}$Dipartimento di Matematica e Informatica dell'Universit\`a di Salerno, 84100 Salerno, Italy}
\address{${}^{3}$ Istituto Nazionale di Fisica Nucleare, Gruppo Collegato di Salerno, 84100 Salerno, Italy}

\address{${}^{4}$Department of Computer Science, Oklahoma State University, Stillwater, OK 74078, USA}

\ead{bruno@sa.infn.it}

%and\\
% Istituto Nazionale di Fisica Nucleare, Gruppo Collegato di Salerno, 84100 Salerno, Italy}

\bigskip

\begin{abstract}
We discuss the presence of a geometrical phase in the evolution of a qubit state and its gauge structure. The time evolution operator is found to be the free energy operator, rather than the Hamiltonian operator.
\end{abstract}

As well known, gauge theories provide important tools in the study of the  dynamics of elementary particles and condensed matter systems. The physics of many systems is indeed described by Lagrangians which are invariant under local gauge transformation groups. The relevance of global geometric properties and gauge invariance has been displayed also in the quantum computing framework \cite{Zanardi:1999,Zanardi:1999b,Wilczek:1984}. In this paper, we report recent results \cite{Bruno:2010} according to which the time evolution of a two level system, such as a  qubit state, is governed by a gauge field structure and geometric phases appear. We construct the covariant derivative operator and show that it is related to free energy.

Let us start by considering the standard orthonormal basis of two vectors $|0\rangle$ and $|1\rangle$, eigenvectors of the operator $H = \omega_1 |0 \rangle \langle 0| + \omega_2 |1 \rangle \langle 1|$ with eigenvalues $\omega_1$ and $\omega_2$, respectively. By use of a rotation operator in the plane $\{ |0\rangle,|1\rangle \}$ one may prepare the states
\begin{eqnarray}\lab{phi}
|\phi  \rangle & = &
\alpha  \;|0\rangle \;+\;  \beta  \; |1\rangle \,,~
\\ \lab{psi}
|\psi  \rangle & = &
-\beta  \;|0\rangle \;+\; \alpha  \; |1\rangle \,.
\end{eqnarray}
The coefficients $\alpha$ and $\beta$ fulfill the relation $|\alpha|^{2} + |\beta|^{2} = 1$. Then, in full generality, we may put $\alpha = e^{i\gamma_1} \cos \theta$ and $\beta = e^{i\gamma_2} \sin \theta $. Here  we assume $\gamma_1 = \gamma_2 = 0$ (in this paper, we do not consider the initialization problem \cite{SK03,SK04,SK05}).

Let $t$ denote the time parameter. $H$ plays the role of the Hamiltonian operator and the time evolution of the states (\ref{phi}) and (\ref{psi}) is %
\bea \lab{tev}
|\phi (t)\rangle &=& e^{-iHt} |\phi (0)\rangle  =
%e^{-i\omega_1 t} \cos \theta |0\rangle + e^{-i\omega_2 t} \sin \theta |1\rangle
%\\ &=&
e^{-i\omega_1 t} (\cos \theta \; |0\rangle + e^{-i(\omega_2-\omega_1)t} \sin \theta \; |1\rangle) ~,\\
\lab{t02}
|\psi (t)\rangle &=& e^{-iHt} |\psi (0)\rangle = e^{-i\omega_1 t} ( -\sin \theta \; |0\rangle \,+\, e^{-i(\omega_2-\omega_1)t} \cos \theta \; |1\rangle)\,.
\eea
with $\langle\phi (t)|\psi (t)\rangle = 0$ and $\langle\psi (t)|\psi (t)\rangle = 1$, $\langle\phi (t)|\phi (t)\rangle = 1$, for all $t$. We note  that the states $|\phi (t)\rangle$ and $|\psi (t)\rangle$, for all $t$ and $\omega_1 \neq \omega_2$, are not eigenstates of $H$. From these states we have
\begin{eqnarray}\label{omee}
&&\langle \phi(t)|\; H\;|\phi(t)\rangle = \omega_{1}\;\cos^2\theta + \omega_{2}\;\sin^2\theta \,= \,\omega_{\phi \phi}  ~,\\ [0.2cm]
\label{ommm}
&&\langle \psi(t)|\; H\;|\psi(t)\rangle = \omega_{1}\;\sin^2\theta + \omega_{2}\;\cos^2\theta \,=\, \omega_{\psi \psi}~,\\
\label{omem}
&&\langle \psi(t)|\; H\;|\phi(t)\rangle = \frac{1}{2} ( \omega_{2}- \omega_{1})\;\sin 2\theta \,= \, \omega_{\phi \psi}~ ,
\end{eqnarray}
and  $\omega_{\phi \psi}=\omega_{\psi \phi}$. Notice that the matrix elements of $H$ in these equations are time-independent. We also have
\begin{eqnarray}\label{tang}
tg2\theta \,=\, \frac{2\omega_{\phi \psi}}{\delta \omega}\,,
\end{eqnarray}
where $\delta \omega \equiv \omega_{\psi\psi }-\omega_{\phi\phi }$. By using Eqs.~(\ref{tev}) and  (\ref{t02}), the operator $H$ can be expressed for any $t$ as
\begin{eqnarray} \label{htr}
H = \om_{\phi \phi} |\phi (t) \rangle \langle \phi (t)| + \om_{\psi  \psi}|\psi (t) \rangle \langle \psi (t)| + \om_{\phi \psi} (|\phi (t) \rangle \langle \psi (t)| + |\psi (t) \rangle \langle \phi (t)|) ~.
\end{eqnarray}
We come now to the discussion of the Berry-like phase involved in the qubit time evolution. The state $|\phi(t)\rangle$, apart from
a phase factor, reproduces the original state $|\phi(0)\rangle$
after a period $T= \frac{2\pi}{\omega_{2} - \omega_{1}}$ :
\begin{equation}\label{nue2}
|\phi(T)\rangle = e^{i {\varphi}} |\phi(0)\rangle ~, \qquad
{\varphi}= - \frac{2\pi \omega_{1}}{\omega_{2} - \omega_{1}} \, ,
\end{equation}
where it is essential that $\omega_1 \neq \omega_2$.
One thus recognizes that such a time evolution does contain a purely
geometric part, i.e. the Berry-like phase. Indeed, we separate the geometric and ``dynamical'' phases following the standard procedure \cite{AA87}:
\begin{eqnarray}\label{ber1}
\beta_{\phi}&=& \varphi + \int_{0}^{T}
\;\langle \phi(t)|\; i\partial_t\;|\phi(t)\rangle \,dt \;= \; 2 \pi \sin^{2}\theta  \, .
\end{eqnarray}
$\beta_{\phi}$ is independent from $\omega_1$ and $\omega_2$ and depends only on the ``mixing angle'' $\theta$. Similarly, for the state $|\psi (t)\rangle$ one finds $\beta_{\psi}\,= 2 \pi \cos^{2}\theta$ and thus $\beta_\psi + \beta_{\phi} = 2\pi$ for any $\theta$.
We can thus rewrite (\ref{nue2}) as
\begin{equation}\label{nue2b}
|\phi(T)\rangle = e^{i 2\pi \sin^2\theta} e^{-i \omega_{\phi \phi} T}
|\phi(0)\rangle \, .
\end{equation}
The meaning of Eqs.~(\ref{ber1})-(\ref{nue2b}) can be better understood by noticing that, for any $t$,
\begin{equation}\label{ovlap1}
\langle \phi(0)|\phi(t)\rangle \,=\,e^{-i \omega_{1} t}
\cos^2\theta  + e^{-i \omega_{2} t} \sin^2\theta \, .
\end{equation}
Thus, as an effect of the non vanishing difference $\omega_{-} = \om_2  - \om_1$ of the  phases, the components $|0 \rangle$ and $|1 \rangle$ evolve with different ``weights'' and  the state $|\phi (t)\rangle$
``rotates'' as shown by Eq.~(\ref{ovlap1}). At $t=T$,
\begin{equation}\label{ovlap2}
\langle \phi(0)|\phi(T)\rangle \,=\,
e^{i\varphi} \,=\, e^{i \beta_\phi} e^{-i \omega_{\phi \phi} T}\, ,
\end{equation}
i.e. $|\phi(T)\rangle$ differs from $|\phi(0)\rangle$ by a phase
$\varphi$, part of which is a geometric ``tilt'' (the Berry-like phase $\beta_{\phi}$) and the
other part is of dynamical origin. This is similar to what happens in the context of particle mixing (see ref. \cite{Blasone:1999}, which we closely follow in our presentation below).
In general, for $t= T +\tau$, we have
\begin{eqnarray}\nonumber
\langle \phi(0)|\phi(t)\rangle &=&
e^{i\varphi} \,  \langle \phi(0)|\phi(\tau)\rangle
\\ \label{ovlap3}
&=& e^{i 2\pi \sin^2\theta} e^{-i \omega_{\phi \phi} T}
\left( e^{-i \omega_{1} \tau} \cos^2\theta  + e^{-i \omega_{2} \tau}
\sin^2\theta \right) \, .
\end{eqnarray}
Also notice that
\begin{equation}\label{ovlap4}
\langle \psi(0)|\phi(t)\rangle \,=\,\frac{1}{2}\,
e^{i\varphi}  e^{-i \omega_{1} \tau}\,\sin
2\theta \left( e^{-i(\omega_2 -\omega_1) \tau} - 1\right), \qquad
{\rm for} \; t= T +\tau\, ,
\end{equation}
which is zero  only at  $t=T$.
Eq.~(\ref{ovlap4}) expresses the fact that $|\phi(t)\rangle$
``oscillates'', getting a component  of the $|\psi(0)\rangle$ state, besides getting the
Berry-like phase. At $t=T$, $|\phi(t)\rangle$ and $|\psi(0)\rangle$ are again each other orthogonal states.

Generalization to $n-$cycles, $n = 1, 2, 3,...$, is also interesting. Eq.~(\ref{ber1}) can be rewritten for the  $n-$cycle case as
\begin{equation}\label{ber1n}
\beta^{(n)}_{\phi}\,= \, \int_{0}^{nT}
\;\langle \phi(t)|\; i\partial_t -\omega_1\;|\phi(t)\rangle \,dt
\, =\, 2 \pi \,n\,\sin^{2}\theta \, ,
\end{equation}
and Eq.~(\ref{ovlap3}) becomes
\begin{equation}\label{ovlapn}
\langle \phi(0)|\phi(t)\rangle \,=\,
e^{i n\varphi} \, \langle \phi(0)|\phi(\tau)\rangle \,, \qquad {\rm for} ~~\; t= n T +\tau \, .
\end{equation}
Similarly, in Eq.~(\ref{ovlap4}) one obtains the phase $e^{i n\varphi}$ instead of
$e^{i \varphi}$.
Eq.~(\ref{ber1n}) shows that the Berry-like phase acts as a ``counter'' of
$|\phi(t)\rangle$ oscillations, adding up $2 \pi \,\sin^{2}\theta$ to the phase
of the $|\phi(t)\rangle$ state after each complete oscillation. We observe that $\beta^{(n)}_{\phi}$ in Eq.~(\ref{ber1n}) can be rewritten as
\begin{equation}\label{ber1n2}
\beta^{(n)}_{\phi}\,= \, \int_{0}^{nT}
\;\langle \phi(t)|\; U^{-1}(t)\,i \partial_t \,
\Big( U(t)\;|\phi(t)\rangle\Big)
\,dt \,= \, \int_{0}^{nT}
 \langle \tilde{\phi}(t)|\;i \partial_t
|\tilde{\phi}(t)\rangle \,dt \, =\, 2 \pi \,n\,\sin^{2}\theta \, ,
\end{equation}
with $U(t)=e^{-i f(t)}$, where $f(t)=f(0) -\omega_1 t$, with $f(0)$ an arbitrary constant, and
\begin{equation}\label{tildenue}
|\tilde{\phi}(t)\rangle \equiv U(t)|\phi(t)\rangle \, = \,
 e^{-i f(0)} \left(\cos\theta\;|0\rangle \;+\;
e^{-i (\omega_{2}-\omega_{1}) t}\; \sin\theta\; |1\rangle \;
\right)\, .
\end{equation}

We can regard $|\phi(t)\rangle \rar U(t)|\phi(t)\rangle  = |\tilde{\phi}(t)\rangle$ as a local (in time) gauge transformation of the state $|\phi(t)\rangle$. In contrast with the state $|\phi(t)\rangle$, the gauge transformed state $|\tilde{\phi}(t)\rangle$
is not ``tilted'' in its time evolution:
\begin{equation}\label{ovlaptilde}
\langle \tilde{\phi}(0)|\tilde{\phi}(t)\rangle \,=\,
\langle \tilde{\phi}(0)|\tilde{\phi}(\tau)\rangle \,,
\qquad \qquad {\rm for} ~~ \; t= n T +\tau \, ,
\end{equation}
which has to be compared with Eq.~(\ref{ovlapn}). From
Eq.~(\ref{tildenue}) we see that time evolution
only affects the $|1\rangle$ component of the state
$|\tilde{\phi}(t)\rangle$. The gauge transformation acts as a ``filter'' freezing out time evolution of the $|0\rangle$ state component, so that we have
\begin{eqnarray}\nonumber
i\partial_t |\tilde{\phi}(t)\rangle &=&
(\omega_2 -\omega_1)  e^{-i f(0)} e^{-i(\omega_2 -\omega_1)t} \sin\theta
|1\rangle
\\ \nonumber
&=& (H -\omega_1) e^{-i f(0)}\left(\cos\theta\;|0\rangle \;+\;
e^{-i (\omega_{2}-\omega_{1}) t}\; \sin\theta\; |1\rangle \;
\right)
\\ \label{patilde}
&=&(H -\omega_1)|\tilde{\phi}(t)\rangle \,,
\end{eqnarray}
namely
\begin{eqnarray} \label{Dtilde}
- i(\partial_t + iH) |\tilde{\phi}(t)\rangle = \omega_1 |\tilde{\phi}(t)\rangle \,.
\end{eqnarray}

Eq.~(\ref{ber1n2}) actually provides
an alternative way for defining the Berry-like phase \cite{AA87},  which
makes use of the  state $|\tilde{\phi}(t)\rangle$ given in
Eq.~(\ref{tildenue}). Eq.~(\ref{ber1n2}) directly gives us the geometric
phase because the quantity $i\langle \tilde{\phi}(t)|(i\pa_{t}|
\tilde{\phi}(t)\rangle\,dt) $ is the overlap of
$|\tilde{\phi}(t)\rangle$ with  its ``parallel transported'' $(i\pa_{t}|
\tilde{\phi}(t)\rangle\,dt)$ at $t+dt$.

Another geometric invariant is the Anandan--Aharonov phase \cite{AA90},  defined as $s= 2 \int \Delta \om (t) dt$.
It has the advantage to be well defined also for systems with non-cyclic evolution. Since $ \Delta \om =  \omega_{\psi \phi}$, we have
\begin{equation}\label{inv1}
s_{n}\, =2\, \int_0^{nT} \,\omega_{\psi \phi} \, dt \,= 2\,  \pi\,n \sin 2\theta
\,.
\end{equation}
Such an invariant represents the distance between qubit evolution states, as measured by the Fubini--Study metric, in the projective Hilbert space
$\mathcal{P}$ \cite{Bruno:2010} (see also ref. \cite{Celeghini:2009}).

We now show the gauge structure underlying the time evolution of two level systems.
The motion equations for the $|\phi (t)\rangle$ and $|\psi (t) \rangle$ are computed by using the operator $H$ in Eq.(\ref{htr})  written in terms of time dependent states. We have $H |\phi(t)\rangle = \,\om_{\phi\phi}\,|\phi (t)\rangle\,+ \,\om_{\phi \psi}\,\,|\psi(t)\rangle$ and $H |\psi(t)\rangle = \,\om_{\psi\psi}\,|\psi (t)\rangle\,+ \,\om_{\phi \psi}\,\,|\phi(t)\rangle$. Thus, in compact form, we have
\begin{eqnarray}\label{HeisenbergComp}
 i\,\partial_{t}\, |\zeta(t)\rangle\,=\,\om_{d}\,|\zeta(t)\rangle\,+\,\om_{\phi \psi}\,\sigma_{1}\,|\zeta(t)\rangle \,,
\end{eqnarray}
where $|\zeta(t)\rangle\,=\,(|\phi(t)\rangle \,,|\psi(t)\rangle )^{T}$ and $\om_{d}=diag(\om_{\phi\phi },\om_{\psi\psi })$.
We use the notation  $g \,\equiv\,\tan 2\theta$ and $\om_{\phi \psi}\,=\,\frac{1}{2} g\, \delta \omega$ from Eq.~(\ref{tang}) and we set  $A_{0} = A_{0}^{(1)}\,\sigma_{1} \,\equiv\,\frac{1}{2} \, \delta \omega \,\sigma_{1}$. The covariant derivative is defined by the following relation
\bea\label{covarDer}
D_{t}\,=\,\partial_{t}\,+\,i\,\om_{\phi \psi}\,\sigma_{1}
\,=\partial_{t}\,+\,i \,g\, A_{0}^{(1)}\,\sigma_{1} \,
%=\,\partial_{t}\,+\,\frac{i}{2} \tan 2\theta\, \delta \omega \,\sigma_{1}\,
\eea
where $g$ behaves as a coupling constant and $A_{0}$ represents the gauge operator field. The motion equation
(\ref{HeisenbergComp}) can be then expressed as
\bea \label{D}
iD_{t}\,|\zeta(t)\rangle\,=\,\om_{d}\,|\zeta(t)\rangle\,.
\eea
We point out that under the following transformations
\bea \label{D1}
&{}& D_{t}'\,=\,\partial_{t}\,+\,i \,g\, (A_{0}^{(1)}\,\sigma_{1} \, + \partial_{t}\,\lambda(t)\,\sigma_{1})\,,\\
&{}& |\zeta'(t)\rangle = e^{-i g \,\lambda(t)\,\sigma_{1} }|\zeta(t)\rangle\,,
\eea
Eq.(\ref{D}) becomes
\bea
iD_{t}'\,|\zeta'(t)\rangle\,=\,\om_{d}\,|\zeta'(t)\rangle\,.
\eea
Thus defining $U(t) \,\equiv\, e^{-i g \,\lambda(t)\,\sigma_{1} }$, we have
\bea
U(t)\,(iD_{t}\,|\zeta(t)\rangle)\,&=&\,iD_{t}'\,U(t)\,|\zeta(t)\rangle\, \\
g \, {A_{0}^{(1)}}' \,\sigma_{1}\, &=& U(t)\,g \,A_{0}^{(1)}\,\sigma_{1}\,U^{-1}(t) \,+\, i\, (\partial_{t}\,U(t))\,U^{-1}(t)\, ,
\eea
as it should be indeed for a gauge field transformation (see  also ref. \cite{Blasone:2010} where neutrino  mixing is studied). The constant $A_{0}$ is the only non-vanishing component of $A_{\mu}$, this implies that the strength field $F_{\mu \nu}$ is identically zero. This is a feature which, for example, occurs  when the gauge potential is a pure gauge (with non-singular gauge functions).
A typical case (with non-integrable phase conditions) is the one of the Aharonov--Bohm effect \cite{Bohm}.

We now express Eqs.~(\ref{HeisenbergComp}) as
\begin{eqnarray}\label{HeisenbergCompH}
 (H\, -\,\om_{\phi \psi}\,\sigma_{1})\,|\zeta(t)\rangle = \,\om_{d}\,|\zeta(t)\rangle\,,
\end{eqnarray}
and introduce the operator F so that
\begin{eqnarray}\label{FreeEn}
 F \,=\, H\, -\,\om_{\phi \psi}\,\sigma_{1} \;.
\end{eqnarray}
Such an operator may be considered as the free energy operator provided that  the term $\om_{\phi \psi}\,\sigma_{1}$ is identified with the entropy term $T S$ in the traditional free energy expression. In such a case we may  put the ``temperature" $T \, = g$ and the entropy $S = A_{0} = \frac{1}{2} \, \delta \omega \,\sigma_{1}$.
The entropy term $T S$, in terms of the time dependent states, is written as
\begin{eqnarray} \label{TS}
 T S=  \om_{\phi \psi} (|\phi (t) \rangle \langle \psi (t)| + |\psi (t) \rangle \langle \phi (t)|) ~.
\end{eqnarray}
Moreover, we note that the geometric invariant $s$ defined in Eq.~(\ref{inv1}) is related to the entropy term; indeed we have
\begin{eqnarray} \label{TSs}
 \int_0^{n T} \langle \zeta (t)| T S \sigma_{1}|\zeta (t)\rangle ~dt = \int_0^{n T} \langle \zeta (t)| g\, A_{0}^{(1)}|\zeta (t)\rangle ~dt =  2\,\int_0^{n T}\om_{\phi \psi} \, dt = s_{n} .
\end{eqnarray}
Thus, the relation between $T S$ and the variance of the energy $\Delta \, \om\,=\,\om_{\phi\psi}$ is induced by the non diagonal elements of $H$.

In conclusion, we have shown the presence of geometric phases in the evolution of a two level system and studied its gauge structure. We have computed the covariant derivative and pointed out that it acts as the free energy with the gauge field acting as the entropy.
In such a picture time evolution is thus controlled  by the free energy. When applied to a qubit state, these results may be of interest in quantum computing studies.

\section*{Acknowledgements}
Partial financial support from INFN is acknowledged.

\medskip
%%%%%%%%%%%%%%%%%%%%%%%%%%%%%%%%%%%%%%%%%%%%%%%%%%%%%%%%%%%%%%%%%%%%%%%%%%%%

\end{document}